\documentclass[aip,jcp]{revtex4-1}

\usepackage{graphicx}%
\usepackage{dcolumn}%
\usepackage{bm}%

\usepackage[font=footnotesize]{caption}

\begin{document}
\title{Multiphoton Photoelectron Circular Dichroism of Limonene with Independent Polarization State Control of the Bound-Bound and Bound-Continuum Transitions}

\author{S. Beaulieu$^{1,2}$,
A. Comby$^{1}$,
D. Descamps$^{1}$,
S. Petit$^{1}$,
F. L\'egar\'e$^{2}$,
B. Fabre$^{1}$,
V. Blanchet$^{1}$, 
Y. Mairesse$^{1}$} 
\bigskip

\affiliation{
$^1$Universit\'e de Bordeaux - CNRS - CEA, CELIA, UMR5107, F33405 Talence, France\\
$^2$ Institut National de la Recherche Scientifique, Centre EMT, J3X1S2, Varennes, Quebec, Canada\\
}

\date{\today}

\begin{abstract}

Photoionization of randomly oriented chiral molecules with circularly polarized light leads to a strong forward/backward asymmetry in the photoelectron angular distribution. This chiroptical effect, referred to as Photoelectron Circular Dichroism (PECD), was shown to take place in all ionization regimes, from single photon to tunnel ionization. In the Resonance Enhanced Multiphoton Ionisation (REMPI) regime, where most of the table-top PECD experiments have been performed, understanding the role of the intermediate resonances is currently the subject of experimental and theoretical investigations. In an attempt to decouple the role of bound-bound and bound-continuum transitions in REMPI-PECD, we photoionized the (+)-limonene enantiomer using two-color laser fields in [1+1'] and [2+2'] ionization schemes, where the polarization state of each color can be controlled independently. We demonstrate that the main effect of the bound-bound transition is to break the sample isotropy by orientation-dependent photoexcitation, in agreement with recent theoretical predictions. We show that the angular distribution of PECD strongly depends on the anisotropy of photoexcitation to the intermediate state, which is different for circularly and linearly polarized laser pulses. On the contrary, the helicity of the pulse that drives the bound-bound transition is shown to have a negligible effect on the PECD. 

\end{abstract}
\maketitle

Photoelectron circular dichroism (PECD) is a chiroptical effect which was theoretically predicted in the 1970's \cite{ritchie76}. Unlike most chiroptical phenomena, which rely on magnetic dipole or electric quadrupole effects, PECD can be fully described within the framework of the electric dipole approximation. As a consequence,  PECD is several orders of magnitude larger than conventional chiroptical effects, which are in general extremely weak and thus difficult to use for studies in gas-phase samples. The first experimental observation of PECD from an ensemble of randomly oriented chiral molecules was performed in 2001, using Vacuum Ultra Violet (VUV) synchrotron radiation \cite{bowering01}. Since then, many studies relying on single photon ionization have been conducted, showing that PECD is a powerful probe of molecular chirality (for a review see \cite{powis08-2}), sensitive to electronic structure \cite{powis08}, vibrational excitation \cite{garcia13,rafiee_fanood_intense_2017}, molecular conformation \cite{garcia08,turchini09,tia14}, structural isomerism \cite{powis08-1,nahon16}, clustering \cite{powis13} as well as chemical substitution \cite{stener04,garcia14}. Lately, it was also demonstrated that single photon PECD experiments could be performed using table-top femtosecond elliptically polarized VUV sources from resonant high-order harmonic generation \cite{ferre15}. 

Recently, a whole new field of PECD studies emerged, relying on the use of UV, visible or IR table-top femtosecond laser sources. In two pioneering experiments, circularly polarized femtosecond UV pulses were used to photoionize enantiopure chiral molecules in a [2+1] Resonant Enhanced Multiphoton Ionisation (REMPI) scheme, and led to the observation of large PECDs, in the 10$\%$ range \cite{lux12,lehmann13}. These results were interpreted as a sequential mechanism, where resonant two-photon transition brings the molecules into low-lying Rydberg states and the third photon subsequently photoionizes the system. Importantly, REMPI-PECD was shown to be highly sensitive to isomerism \cite{lux12} and vibrational excitation \cite{fanood16,beaulieu16-1}. PECD was also observed in other ionization regimes: low order \cite{lux16} and high order \cite{beaulieu16-1} above-threshold ionization, and even in the tunnel ionization regime \cite{beaulieu16-1}. PECD is thus a universal effect, which is inherent to the photoionization of chiral molecules with circularly polarized light, in all ionization regimes. Very recently, PECD was extended to ultrafast dynamical studies of photoexcited chiral molecules by femtosecond time-resolved PECD measurements, which established the high sensitivity of this observable to subtle molecular dynamics \cite{comby16,beaulieu16-2}. A self-referenced attosecond photoelectron interferometry has also allowed measuring the temporal profile of the forward and backward electron wavepackets emitted upon photoionization of camphor by circularly polarized laser pulses to investigate the enantiosensitivity of photoionization time delays \cite{beaulieu17}.

REMPI-PECD is of particular importance because it produces very strong signals that can be used to probe molecular dynamics or to quantitatively determine enantiomeric excesses with high accuracy \cite{fanood15,kastner16,greenwood17}. The exact influence of intermediate resonant states in PECD is, however, still an open question. In a very recent high-resolution PECD experiment \cite{kastner17}, Kastner \textit{et al.} have shown a sign switch of PECD depending on which electronic state is used as the intermediate resonance. From a theoretical point of view, a first investigation \cite{dreissigacker14} showed that using the Coulomb-corrected Strong Field Approximation (ccSFA) to treat nonperturbatively the ionization was sufficient to yield a nonzero PECD. Since this approximation intrinsically neglects the intermediate excited states, it is a demonstration that multiphoton PECD can exist in a purely non-resonant case. However, the qualitative agreement between the calculated and experimental PECD was rather poor, showing opposite signs. This could be due to the fact that resonances may change drastically the initial state from which electron scatters before being emitted. To solve this issue, Goetz \textit{et al.} recently developed a complete theoretical framework which combines \textit{ab initio} calculations for the bound-bound transition and perturbation theory for subsequent single photon ionization from the exited state \cite{goetz17}. This theoretical treatment in which the role of bound-bound and bound-continuum transitions in REMPI-PECD are totally decoupled, showed that the primary role of the bound-bound transition is to break the isotropy in the molecular ensemble by creating an anisotropic distribution of photoexcited molecules, due to the orientation-dependence of the transition matrix elements. They demonstrated that the contribution of high-order odd Legendre coefficients ($b_i$, where $i>1$) in a [2+1] ionization scheme was the signature of this anisotropy, and that the ratio between the different odd Legendre coefficients was determined by the electronic character of the intermediate excited state. Finally, they predicted that using circularly polarized photons to drive the bound-bound transition was not necessary to produce PECD, as was also recently demonstrated in a time-dependent PECD experiment in fenchone \cite{comby16}. 

In this paper, motivated by the theoretical work of Goetz \textit{et al.}, we use two-color laser fields with various polarization combinations to experimentally disentangle the role of bound-bound and bound-continuum transition in [1+1'] and [2+2'] REMPI-PECD of (+)-1-Methyl-4-(1-methylethenyl)-cyclohexene (referred to as (+)-limonene). These results are compared to the standard single-color [2+1] scheme, where all photons are circularly polarized (with the same helicity). Our results confirm that the bound-bound transitions do not need to be driven with circularly polarized photons to observe PECD. In addition, we show that the left or right helicity of the exciting photons hardly affects the measured PECD. This demonstrates that the main influence of the photoexcitation step is to create an anisotropy in the sample. This anisotropy depends on the linear or circular nature of the driving field.   

The experiments were performed using the home-built Ti:Sapphire Aurore laser system at CELIA, which can deliver up to 8 mJ energy per pulse centered around 800 nm, at 1 kHz repetition rate, with pulse duration of 25 fs. For the [2+2'] (396 nm + 800 nm) experiments, the pulses were sent in a (50$\%$/50$\%$) Mach-Zehnder interferometer, where one of the two arms was frequency-doubled in a 200 $\mu$m thick type I BBO crystal, with a resulting Full Width Half Maximum (FWHM) bandwidth of 8 nm. In each arm, we placed a broadband zero order quarter wave plate to control the polarization states of the 800 nm and the 396 nm pulses independently. For this setup, after the quarter wave plates, all the reflective optical components were at $<$5$^{o}$ incidence angle, to minimize polarization state artifacts. In the past, we have conducted single-color PECD measurements using either the 400 nm reflected or the 800 nm transmitted beam, and found a very good mirroring between opposite enantiomers. This indicates that the influence of the non-zero incidence on the mirrors is negligible. The pulses were recombined by a dichroic mirror and were focused by using a 1000 mm focal lens into a velocity map imaging spectrometer. The (+)-limonene molecules were introduced as a continuous flow through a 200 $\mu$m nozzle heated at 90$^o$C and located 7 cm away from the laser focus. The pressure in the interaction chamber was typically $1\times{10}^{-6}$ mbar, with a background pressure of $7\times{10}^{-8}$ mbar. For the [1+1'] (199 nm + 402 nm) experiments, the setup was slightly different. One part of the laser beam was used to generate the 199 nm pulses by sum frequency generation between the fundamental 800 nm and its third harmonic (267 nm) in a 100 $\mu$m thick BBO crystal with a resulting FWHM bandwidth of 1.5 nm. The second arm was frequency doubled from 800 nm to 402 nm in a 100 $\mu$m BBO crystal with a resulting FWHM bandwidth of 14 nm. In this setup, the pump and probe pulses are focused in a non-collinear geometry, with a 7$^{o}$ angle between the two beams (\textit{i.e.} without the need of using a recombining dichroic mirror). The pulses, with respective energies of 3$\mu$J and 5$\mu$J, were focused by a 250 mm and a 600 mm lens, respectively, into a pulsed (1 kHz) molecular beam produced by a Even-Lavie valve \cite{even00} with a 250 $\mu$m conical nozzle, at a temperature of 70$^o$C and backed with a carrier gas of helium to avoid cluster formation. No trace of dimer was observed on the mass spectrum or on the ion images. In our experimental conditions we expect the first two conformers of limonene to be present \cite{moreno_conformational_2013}. These conformers only differ by a few kJ/mol and are not distinguished in PES and PECD measurements in one-photon ionization close to the threshold ($\textless$ 10 eV) \cite{rafiee_fanood_intense_2017}. 

In all experiments, the photoelectrons were imaged by an electrostatic lens onto a set of two micro-channel plates (MCP) coupled to a phosphor screen (P43), and recorded by a cooled 12 bit CCD camera. The PECD was measured by recording the photoelectron spectra with left (LCP, $p=+1$) and right (RCP, $p=-1$) circularly polarized laser pulses. In order to remove the influence of slow drifts in the experiments, the polarization state was switched every 10 seconds. Typically, $\sim$3x10$^5$ laser shots for the one color experiment and $\sim$1x10$^6$ for the two color experiment were accumulated to obtain the LCP or RCP images. Measurements using different polarization states were conducted at constant pulse energies. In two-color measurements, the pulse energies were kept low enough to observe ionization only when the two pulses were time-overlapped. We present data recorded only on the enantiomer (+)-R-limonene that has been purchased from Sigma-Aldrich with a purity of 97 $\%$ and an enantiomeric excess at 98 $\%$. 

\begin{figure}
\begin{center}
\includegraphics[width=6cm,keepaspectratio=true]{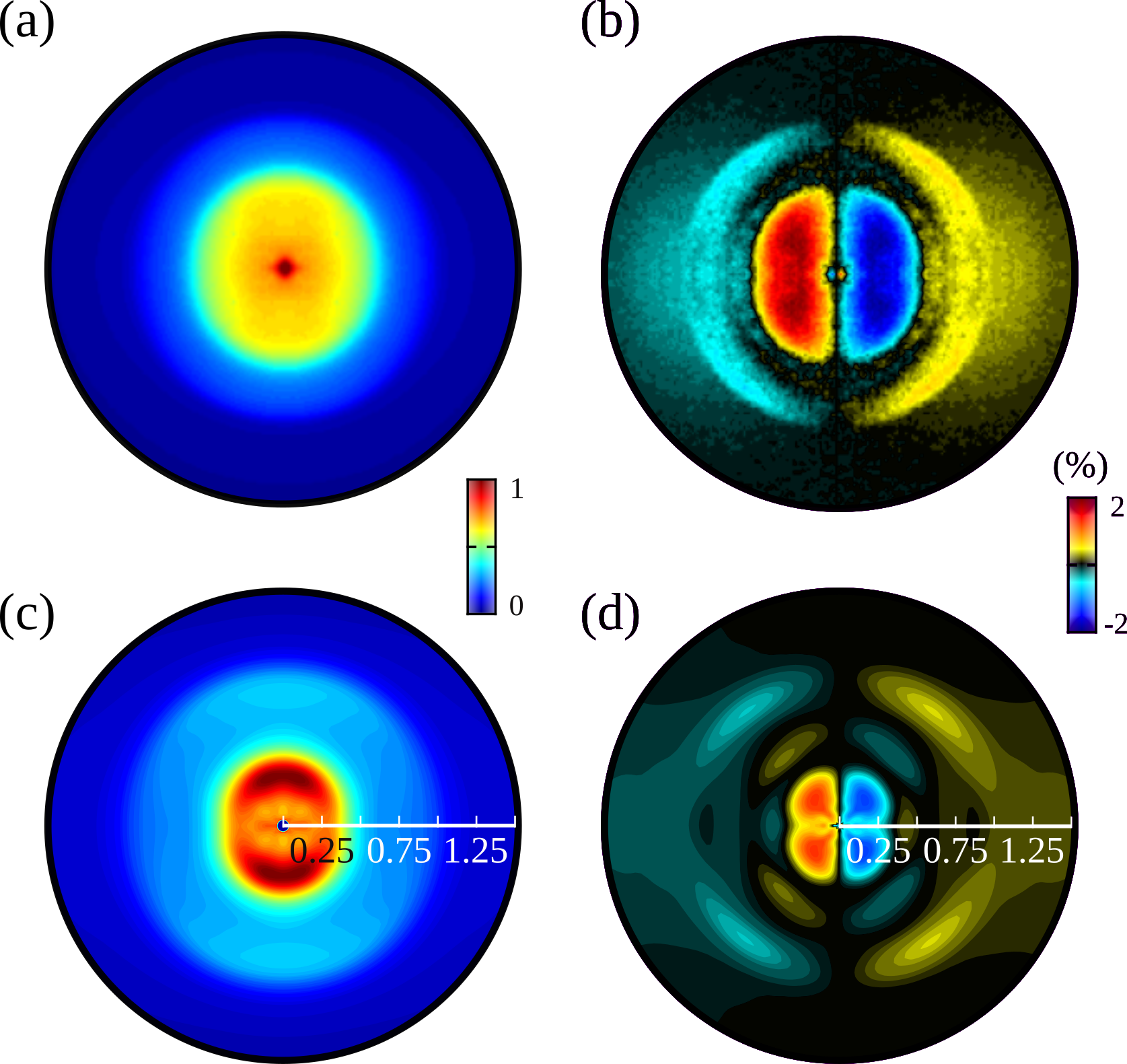}
\caption{Comparison between the raw velocity map image of the photoelectron distribution and a slice through the 3D photoelectron distribution retrieved by the pBASEX method, in (+)-limonene using a [2+1] photoionization scheme at 396 nm. (a) and (b) are the raw velocity map images of the photoelectron distribution and the associated PECD, respectively. In (c) and (d) are slices through the 3D photoelectron distribution retrieved by the pBASEX method and the associated PECD, respectively; the radius represents the energy of the photoelectrons, in eV. The laser propagation axis is horizontal.}
\label{new_raw}
\end{center}
\end{figure}

The velocity map imaging spectrometer collects the two-dimensional projections of three-dimensional photoelectron angular distributions (PAD). The PAD ${S}_{p}(E,\theta )$ resulting from the photoabsorption of $N$ circularly polarized photons can be decomposed as a sum of $2N$ Legendre polynomials $P_i^0$: ${S}_{p}(E,\theta )={\sum }_{i=0}^{2N}{b}_{i}^{p}(E){P}_{i}^0(\cos (\theta ))$, where $p=\pm 1$ is the helicity of the ionizing photons, $E$ is the photoelectron energy and $\theta$ its ejection angle with respect to the light propagation axis. The even coefficients ${b}_{i=2n}^{p}$ fulfill ${b}_{i=2n}^{+1}={b}_{i=2n}^{-1}$, and the odd coefficients ${b}_{i=2n+1}^{+1}=-{b}_{i=2n+1}^{-1}$. We retrieve the even and odd coefficients by fitting respectively the sum (which is left-right and up-down symmetrized) and difference (which is left-right anti-symmetrized and up-down symmetrized) of the images recorded with two helicities, using pBasex \cite{garcia04} (see Fig. \ref{new_raw}). For quantitative analysis we rely on angle-integrated data, namely the angle-averaged photoelectron spectrum $PES=b_0(E)$ and the multiphoton PECD, which is defined as $MPPECD=(2{b}_{1}(E)-\frac{1}{2}{b}_{3}(E)+\frac{1}{4}{b}_{5})/{b}_{0}(E)$ and which is equal to twice the relative difference between the number of electrons emitted in the forward and backward hemispheres for the left helicity \cite{lehmann13}. Note that the images recorded by the CCD camera are binned (2x2) before being processed by pBasex. This does not affect the spectral resolution of our measurements, which is imposed by the velocity map imaging rather than by the discrete acquisition. We estimate this resolution to 50 meV at 100 meV, while our energy sampling after binning is 10 meV in that range, and reaches 40 meV at 1 eV.

We start by investigating the one-color [2+1] REMPI ionization in (+)-(R)-limonene, where all photons are circularly polarized (with the same helicity), as a reference to compare with the two-color [1+1'] and [2+2'] cases, where different combinations of linearly and circularly polarized photons will be used. The one-color [2+1] scenario was previously studied with 420 nm pulses by Rafiee Fanood \textit{et al.} \cite{fanood152}. They reported a PES with one broad peak centered around 0.1 eV assigned  to a 2-photons transition from the HOMO $2\pi$ orbital ($\nu =0$) to the $3s$($\nu =1$) Rydberg state, followed by a single photon ionization to the $2\pi ^{-1}$($\nu =1$) cationic state ($\Delta \nu = 0$ propensity rule for the ionization of Rydberg states) with $\nu$ being the vibrational quantum for C$=$C ring stretching (1447cm$^{-1}$). They observed a maximum PECD of -4$\%$ for this channel (at 0.1 eV). Later, this [2+1] REMPI study was extended to higher energy photons and revealed that two distinct ionization pathways are opened for sufficiently short wavelengths (392 and 396 nm, respectively): one producing a vibrationally and electronically excited cation ($1\pi ^{-1}$($\nu \neq 0$)) and the other one a vibrationless ground state cation ($2\pi ^{-1}$($\nu = 0$)) \cite{beaulieu16-1,fanood16}. The PECD of these two dominant ionization pathways were found to have opposite signs.

\begin{figure}
\begin{center}
\includegraphics[width=9cm,keepaspectratio=true]{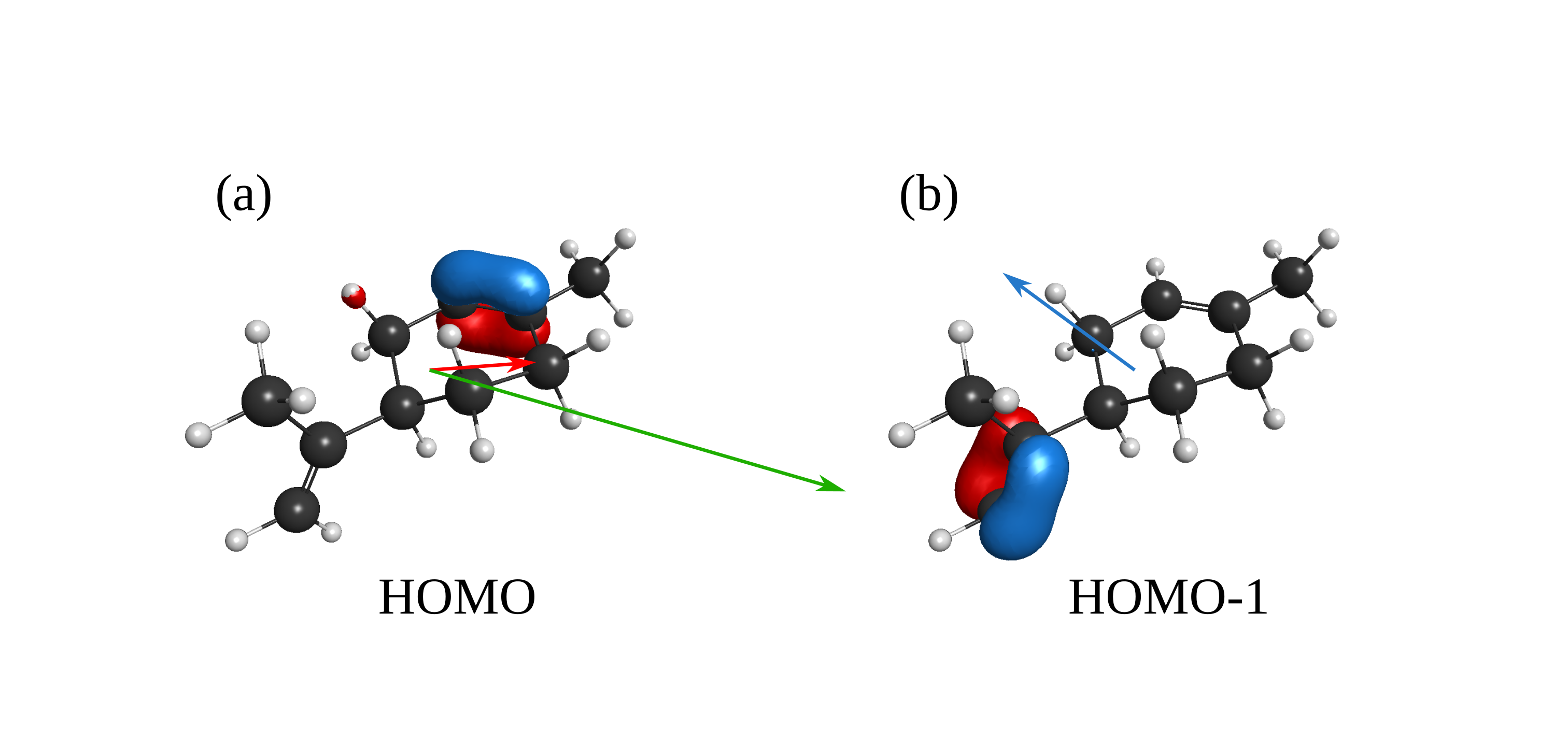}
\caption{Equilibrium geometry of the most stable conformer of (+)-limonene in gas phase at room temperature as well as the representation of the (a) HOMO (2$\pi$) and (b) HOMO-1 (1$\pi$). The red (blue) arrow represents the one-photon transition dipole moment when a HOMO (HOMO-1) hole is created to excite the first accessible low-lying Rydberg orbitale 3s, while the green arrow represents the one-photon transition dipole moment when a HOMO hole is created to excite the second accessible low-lying Rydberg 3p. The oscillator strengths of these transitions are repsectively 0.005 (red), 0.026 (blue) and 0.065 (green) a.u.. }
\label{fig1}
\end{center}
\end{figure}

In the present paper we address the impact of  intermediate resonances reached at energy around 6.23-6.26 eV (2 x 396 nm or 199 nm), where three Rydberg states are energetically accessible from the ground state: 3s$(2\pi)^{-1}$ (calculated electronic energy: 5.79 eV),  3s$(1\pi)^{-1}$ (calculated electronic energy: 6.20 eV) and 3p$(2\pi)^{-1}$ (calculated electronic energy: 6.24 eV) \cite{fanood152}. These calculations are all in good agreement with the CD and UV absorption spectra. In the work presented here, the [2+1], [1+1'] and [2+2'] ionization schemes all lead to a total energy of 9.3-9.4 eV transfer to the system, and thus can be compared to PECD recorded from a totally isotropic sample as one obtained by one-photon VUV ionization around $\sim 9.5$ eV \cite{fanood16,rafiee_fanood_intense_2017}. 

Figure \ref{fig1} shows the HOMO (2$\pi$) and HOMO-1 (1$\pi$) orbitals of (+)-limonene and the one-photon transition dipole moments from the ground electronic state to their two first accessible Rydberg states (3s and 3p) by creating a 2$\pi$ and 1$\pi$ hole in the ion core respectively. The calculations were done using the down-I conformer which is the most abundant at room temperature as well as at several Kelvins \cite{jansik06}. All ab-initio calculations were performed using the GAMESS-US software \cite{schmidt93}. First, the ground state geometry was optimized using density functional theory (DFT) with CAMB3LYP functional and 6-311++G(d,p) basis. Then the energies and transition dipoles of the first 20 excited states of limonene were computed using Time-Dependent density functional theory (TDDFT) with the same functional and basis functions. The choice of this method and basis set are mainly dictated by the compromise between accuracy and computational time. We have compared the calculated energies and the oscillator strengths to the ones of \cite{fanood16} in order to confirm the validity of our choice. One can notice that the HOMO ($2\pi$) orbital is located on the C=C double bond within the cyclohexene ring while the HOMO-1 ($1\pi$) is located on C=C double bond of the isopropenyl moiety. Their respective transition dipole moments to the first accessible Rydberg state are very different, both in amplitude and in direction. This means that a creation of a hole in the HOMO or in the HOMO-1 to populate the first Rydberg states will not photoselect the same subset of molecules within the randomly oriented ensemble. For the HOMO, the transition probability to the 3p state is one order of magnitude larger than to the 3s state (see Fig. \ref{fig1}).

\begin{figure}
\begin{center}
\includegraphics[width=8cm,keepaspectratio=true]{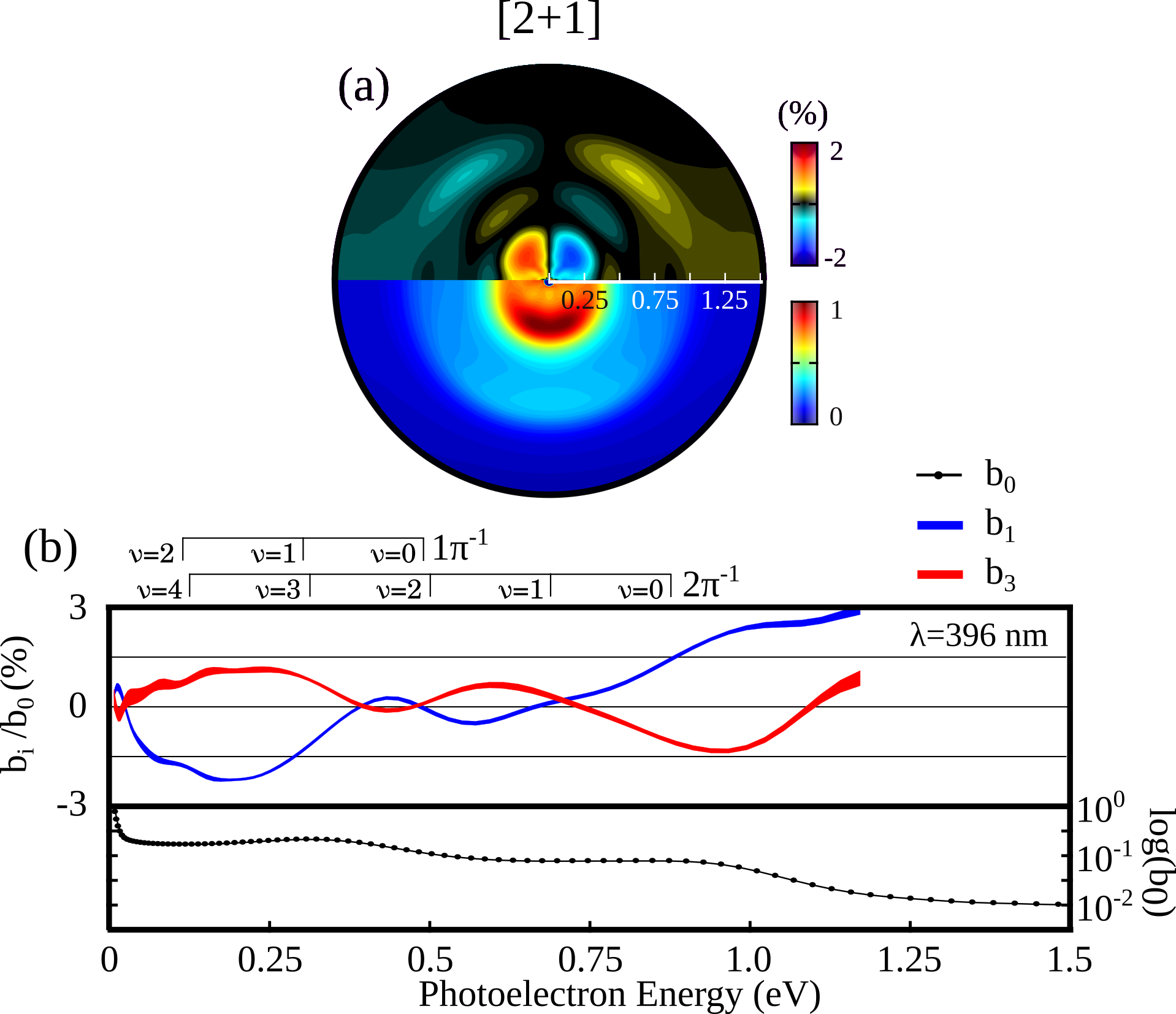}
\caption{PECD of (+)-limonene using a [2+1] photoionization scheme at 396 nm. In (a), the lower hemisphere is a cut of the 3D-PAD (pBasex fit), in linear scale. The upper hemisphere represents the a cut of the 3D-PECD (pBasex fit) that has been further normalized by the absolute maximum of the PES signal. The radius represents the energy of the photoelectrons, in eV.  The laser propagation axis is horizontal. In (b), the upper panel shows the normalized $b_i(E)/b_0(E)$ (i=1,3) odd Legendre coefficients. The thickness of the lines ($b_i(E)/b_0(E)$) represent the 90 $\%$ confidence interval determined using Student's statistics, based on 25 individual measurements of $b_i(E)/b_0(E)$ (each individual measurement is an average over 10000 laser shots, for each helicity). The PES $b_0(E)$ is shown in logarithmic scale in the lower panel. Above (b) is a spectroscopic identification of the vibrational progression for the population of cationic ground- ($2\pi^{-1}$) and excited-state ($1\pi^{-1}$). As in  \cite{fanood16} \cite{lehmann13} \cite{smialek12}, we assume a vibrational excitation of the ring C=C stretching mode ($\sim$ 1500 $cm^{-1}$ = 185 meV). 
}
\label{fig2}
\end{center}
\end{figure}

Figure \ref{fig2} shows the one-color [2+1] photoionization of (+)-limonene molecules with 396 nm femtosecond pulses (3.13 eV photon energy, 9.39 eV for the 3-photon absorption). The PES presents two main peaks centered around 0.3 eV and 0.9 eV. The first adiabatic ionization potential of limonene is 8.505$\pm$0.05 eV \cite{rafiee_fanood_intense_2017}, and the the first electronically excited cationic state ($1\pi ^{-1}$) has a calculated vertical energy of 8.9 eV \cite{fanood16}. The peak around 0.9 eV is thus assigned to ionization of the HOMO leaving the ion in the vibrational ground state ($2\pi^{-1}(\nu=0)$). This assignment is the same as in a previous studies \cite{fanood16}. Due to the $\Delta \nu=0$ ionization rule for Rydberg excitation, the kinetic energy of the photoelectron is expected to be close to $\hbar \nu_{probe}-E_b$, where $E_b$ is the binding energy of the Rydberg state. The 0.9 eV peak is thus associated to an intermediate resonance with the 3p$(2\pi)^{-1} (\nu=0)$ state reached by two-photon absorption. This is consistent with our calculation (see Fig. \ref{fig1}). The assignment of the second peak, centered around 0.3 eV, is more challenging. Based on energetic consideration, this peak can come either from the production of a vibrationally excited ground state cation ($2\pi ^{-1}$($\nu = 3$)) or by the production of a cation in its first vibrational and electronic excited states ($1\pi ^{-1}$($\nu = 1$)). Recently, a clever comparison between VUV (9.5 eV) and REMPI (3x392 nm) PES of limonene has allowed assigning this low-energy peak to the formation of the electronically excited cationic state  \cite{fanood16}. The associated two-photon resonance is the 3s$(1\pi)^{-1} (\nu=1)$ state. 

We now focus on the PECD of (+)- limonene in the one-color [2+1] photoionization scheme shown in Fig.\ref{fig2}. All the fitted coefficients (in Legendre polynomials decomposition) presented in the paper are listed in a synthetic Table \ref{tab1}. In the upper hemisphere of Fig. \ref{fig2}(a), the differential angular distribution $({S}_{+1}-{S}_{-1})$ has been normalized by the absolute maximum of the PES $b_0(E)$. This normalization of the image avoids spuriously large values where small amount of photoelectron is produced. However, in order to give quantitative physical quantities that can be compared to previous literature, we have normalized the extracted odd Legendre coefficients by the energy-resolved PES ($b_i(E)/b_0(E)$). Since the number of ionizing photons is $N=3$, Legendre polynomials up to the $5^{th}$ order ($b_5$) can contribute to the PECD. pBasex analysis leads to weak $b_5$ values ($b_5/b_0<$ 0.1 $\%$), which are negligible compared to $b_1/b_0$ and $b_3/b_0$. Consequently, we do not plot $b_5/b_0$ for sake of clarity. The energy dependency of $b_3/b_0$ shows three peaks at $\sim 0.90$, 0.60 and 0.25 eV, maximizing around $\theta=45^{o}$ as expected when the contribution of $b_3$ is significant and when its sign is opposite to the one of $b_1$. While the PECD peaks around 0.9 eV and 0.25 eV clearly correlate with the PES channels identified above ($2\pi^{-1}(\nu=0)$) and $1\pi ^{-1}$($\nu = 1$), respectively), the PECD peak around 0.6 eV is not associated with a clear feature in the PES.  This PECD peak is associated with the production of a cation in the $2\pi^{-1}(\nu=2)$ state,  as assigned in a previous study \cite{beaulieu16-1}. At 0.9 eV, both $b_1$ and $b_3$ have an opposite signs compared with the peaks at 0.25 eV and 0.6 eV. These results thus indicate sign changes of the REMPI-PECD upon vibrational excitation of the ion, and between adjacent molecular orbitals, two properties that are often observed in single-photon PECD measurements \cite{nahon06,powis08,daly11,garcia14}, including in VUV photoionization of limonene \cite{fanood16, rafiee_fanood_intense_2017}. Note that the thickness of the $b_i(E)/b_0(E)$ lines represents the 90 $\%$ confidence interval determined by performing the pBasex analysis for each of the 25 individual PES and PECDs, and by a statistical analysis on the resulting Legendre coefficients. The statistics being similar in all our measurements, we only include the error bars on this figure. 

\begin{figure}
\begin{center}
\includegraphics[width=8 cm,keepaspectratio=true]{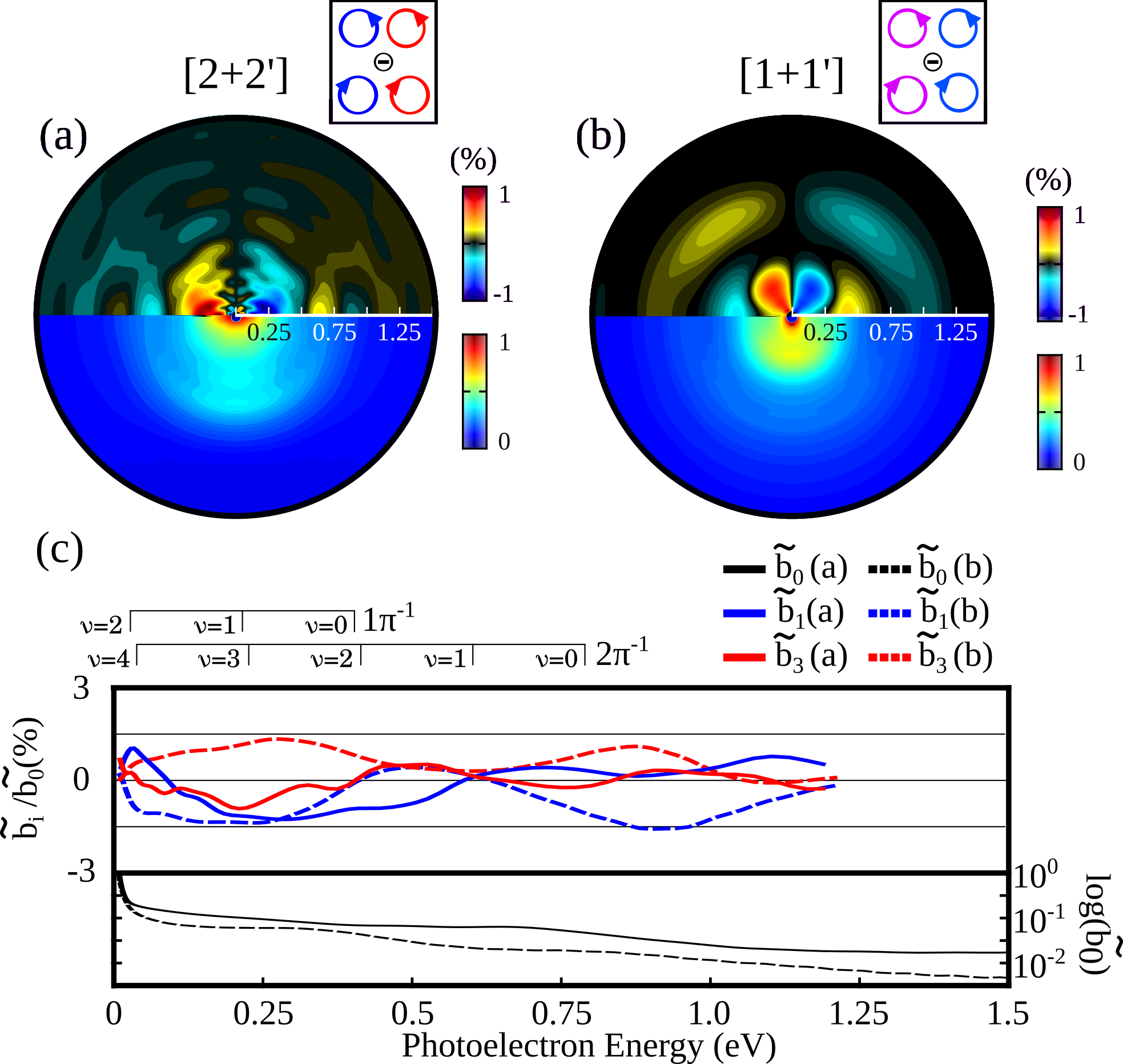}
\caption{PECD of (+)-limonene using corotating two-color laser fields. In (a), we used a [2+2'] (396 nm + 800 nm) scheme while in (b), we used a [1+1'] (199 nm + 402 nm) ionization scheme. As in figure \ref{fig2}, the lower hemispheres are cuts of the 3D-PAD (pBasex fit), in linear scale and the upper hemispheres represent the cuts of the 3D-PECD (pBasex fit) that have been further normalized by the absolute maximum of the PES signal. The radius represents the energy of the photoelectrons, in eV. The laser propagation axis is horizontal. In (c), the upper panel show the normalized polynomials decomposition of the PECD. The thick (dashed) blue line represent the $b_1(E)$ for the [2+2'] ([1+1']) scheme, while the thick (dashed) red line represent the $b_3(E)$ for the [2+2'] ([1+1']) scheme. For the lower panel, the thick (dashed) line represent the PES for the [2+2'] ([1+1']) ionization scheme.  }
\label{fig3}
\end{center}
\end{figure}

Next, we modify the REMPI scheme by keeping the photoexcitation step identical (2 photons at 396 nm) but replacing the photoionization step by a two-photon transition (2 photons at 800 nm), which result in a [2+2'] instead of a [2+1] ionization scheme. Absorption of one 800 nm photon from the low-lying 3s-3p Rydberg states leads to an energy range where many high-lying Rydberg states can be populated. This scheme can thus be considered as a [2+1'+1'] scenario (Fig. \ref{fig3}(a)). Since the 800 and 396 nm pulses are synchronized, this scheme is in competition with others, such as for instance [1+2'+1]. We expect the [2+1'+1'] pathway to be dominant because of its doubly resonant character. The PAD and PECD obtained, in this case, are very different to those from the [2+1] case (Fig. \ref{fig3}(a)). The PES does not show any clear sharp feature, excepted a shoulder at $\sim$ 0.17 eV and a bump around 0.65 eV. The PECD peaks around 0.25 eV with negative $b_1$ and $b_3$ values, both weaker than in the [2+1] scheme (Fig. \ref{fig3}(a)). It shows a complicated angular and energy dependence. We interpret this as the result of the large number of Rydberg states that can be populated around 7.75 eV by the 2+1' transition \cite{smialek12}. PECD is known to be highly sensitive to molecular orbitals and vibrational excitation. Here, many different electronic and vibrational excited states contribute to the signal, with possibly different signs, energy dependences and angular distributions of their PECD. As a result, the total PECD, which is the sum of the PECD from different excited states, is weak and strongly modulated. 

In order to investigate the influence of the photoexcitation pathway in REMPI-PECD, we first replace the two-photon excitation by a single (199 nm) photon excitation, while driving the bound-continuum transition with a single (402 nm) photon. 

For the low energy peak, the PECD has the same sign and angular dependency as in the [2+1] scheme and its magnitude is a bit weaker. The PECD peak around 0.6 eV coming from the $2\pi^{-1}$($\nu =2$) channel present in the one-color [2+1] scheme is absent in the [1+1'] scheme. 
Note that 199 nm excitation lays 30 meV below the 2$\times$396 nm excitation. However, since the laser bandwidth of the 199 nm is $\sim$ 100 meV, the excited Rydberg states would be similar in both schemes. For the high energy peak ($2\pi^{-1}$($\nu =0$)), the sign of the PECD reverses compared to the [2+1] case, while the angular dependency stays similar. These results reveal a significant influence of the excitation scheme on the PECD. They may seem surprising at first sight, because we do not expect selection rules to lead to the excitation of different Rydberg states in such a large polyatomic molecule. However, another effect is at play: the orientation-dependence of the photoexcitation process. This dependence breaks the isotropy of the sample, leading to a partial alignment of the electronically excited molecular ensemble in the polarization plane \cite{lehmann13,artemyev15}. Recent measurements of single-photon ionization of chiral molecules in the molecular frame showed that PECD could have opposite sign for molecules aligned parallel or perpendicular to the light propagation direction, which considerably reduces the value of PECD after averaging over all molecular orientations \cite{tia17}. This effect is consistent with our observations. The two-photon excitation of the [2+1] leads to a sharper alignment distribution than the single-photon process of the [1+1']. The overall value of the PECD is larger in the two-photon excitation case, indicating a stronger PECD from the selected molecules. Furthermore, since PECD can change sign depending on the orientation, different alignment distributions can lead to different signs of PECD, as observed in the high-energy peak, around 0.9 eV. 

The number of photons involved in the photoexcitation and photoionization transitions is not the only degree of freedom in our experiment. We can further decouple the role of bound-bound and bound-continuum transitions by changing the relative polarization state of the pulses. Indeed, the calculations from Goetz \textit{et al.} \cite{goetz17} recently demonstrated that driving the bound-bound transition with linearly polarized photons ($p=0$) followed by photoionization with a circularly polarized photon ($p=\pm1$) should be sufficient to observe PECD in chiral molecules. Contrariwise, they also showed that circular bound-bound excitation of a single state followed by linear photoionization should lead to a fully symmetric PAD with respect to the laser propagation axis (\textit{i.e.} no PECD). 

\begin{figure}
\begin{center}
\includegraphics[width=8 cm,keepaspectratio=true]{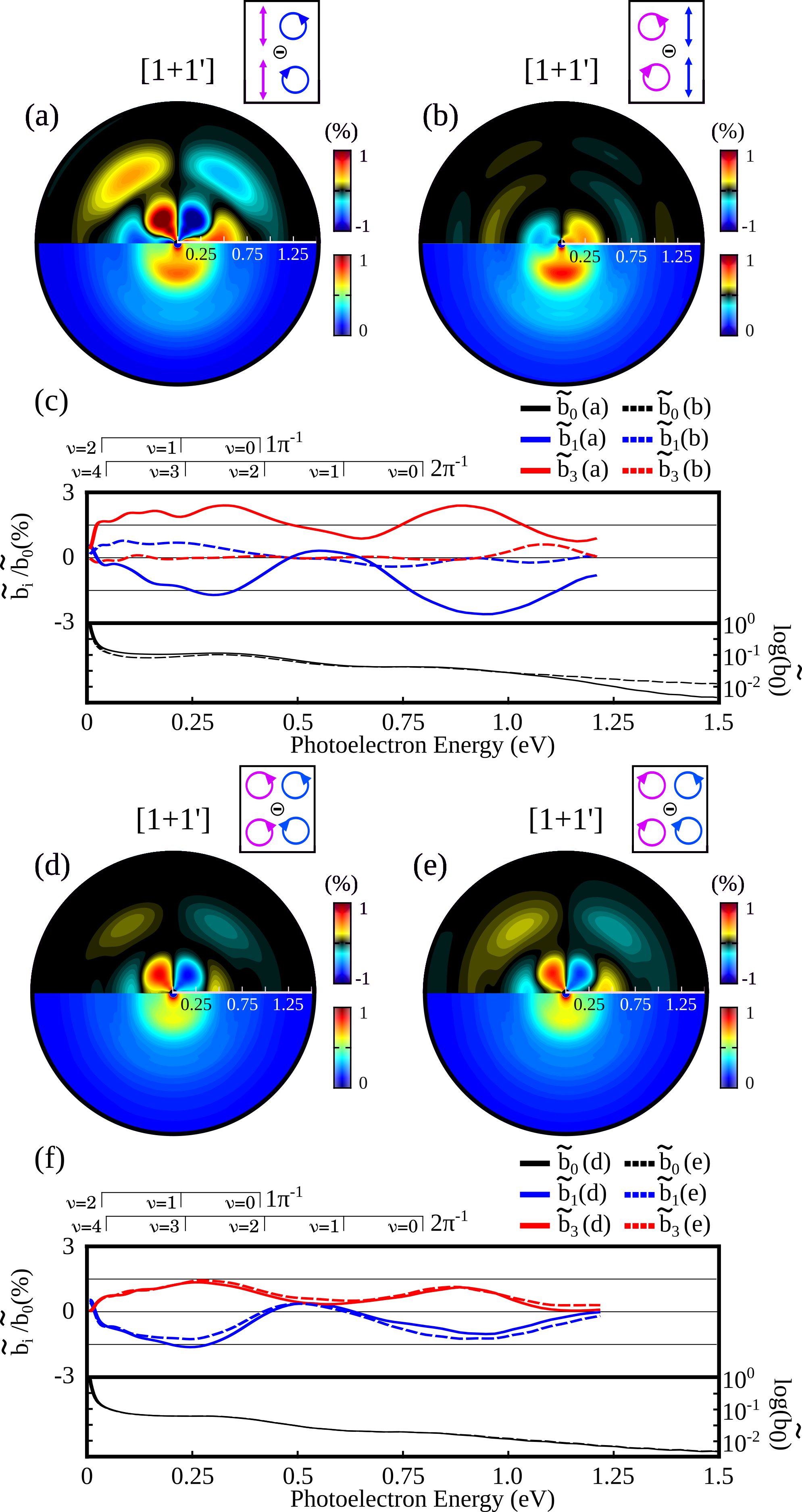}
\caption{PECD of (+)-limonene in [1+1'] (199 nm + 402 nm) ionization scheme using various polarization combinations. As in previous figures \ref{fig2} \ref{fig3}, the lower hemispheres are cuts of the 3D-PAD (pBasex fit), in linear scale and the upper hemispheres represent the cuts of the 3D-PECD (pBasex fit) that have been further normalized by the absolute maximum of the PES signal. In (a), the 199 nm was linearly polarized while the 402 nm was circularly polarized. Note that the polarization axis of the linear beam is always parallel to the detection plane. In (b), the 199 nm pulse was circularly polarized while the 402 nm was linearly polarized. The radius represents the energy of the photoelectrons, in eV. In (c), the upper panel represents the normalized odd Legendre polynomials decomposition of the PECD for (a) (thick lines) and (b) (dashed lines) case. The lower panel represents the PES for (a) (thick black line) and (b) (dashed black line) case. For (d) and (e), we have done the same analysis but using different polarization scheme. As in (a) and (b), the polarization of both 199 nm and 402 nm are indicated in the square black box. The purple arrow represents the polarization state of the 199 nm pulse while the blue arrow represents the polarization state of the 402 nm pulse. The laser propagation axis is horizontal. }
\label{fig4}
\end{center}
\end{figure}

\begin{table*}[t]
  \centering
 \scalebox{0.8}{
\begin{tabular}{| c | c || c | c | c | c | c | c | c |}
\hline
\multicolumn{2}{|c||}{Ionisation Scheme}& [2+1] & [1+1'] & [1+1'] & [1+1']  & [1+1'] & [1+1']  & [1] \\ \hline
\multicolumn{2}{|c||}{Polarization State} & (1,1)-(-1,-1) & (1,1)-(-1,-1) & (0,1)-(0,-1) & (1,0)-(-1,0) & (1,1)-(1,-1)  & (-1,1)-(-1,-1)&\\ \hline 
\multicolumn{2}{|c||}{Figure} & Fig \ref{fig2} (a)  & Fig \ref{fig3} (b) & Fig \ref{fig4} (a) & Fig \ref{fig4} (b) & Fig \ref{fig4} (c) & Fig \ref{fig4} (d) & ref. \cite{rafiee_fanood_intense_2017} \\ \hline \hline
 $2\pi^{-1}$ & $b_1 / b_0$ (0.90 eV) & 1.7 \%  & -1.6 \% & -2.5 \%& 0 \%  & -1.0 \% & -1.2 \% & -1.0 \% \\ \hline
 $2\pi^{-1}$ & $b_3 / b_0$ (0.90 eV) & -1.2 \% & 1.0 \%  & 2.5 \% & 0 \%  & 1.0 \%  & 1.0 \% & - \\ \hline
$2\pi^{-1}$ & $MPPECD$ (0.90 eV) & 4.0 \% & -3.7 \%  & -6.3\% & 0 \% & -2.5 \%  & -2.9 \%  & -2.0 \% \\ \hline
 $1\pi^{-1}$ & $b_1 / b_0$ (0.30 eV) & -1.3 \% & -1.1 \% & -1.8 \% & 0.5 \% & -1.5 \% & -1.1 \%  & -2.0 \% \\ \hline
 $1\pi^{-1}$ & $b_3 / b_0$ (0.30 eV) & 0.9 \% & 1.4 \% & 2.5 \% & 0 \%  & 1.3 \% & 1.4 \% & -  \\ \hline
$1\pi^{-1}$ & $MPPECD$ (0.30 eV)  & -3.1 \% &  -2.9 \%  &  -4.9 \% & 1.0 \%  & -3.7 \%  & -2.9 \%  &  -4.0 \% \\ \hline
\end{tabular}}
  \caption{Odd Legendre polynomial coefficients decomposition of the forward-backward asymmetry for the different photoionization and polarization schemes in R-(+) Limonene. For the polarization state description, we used the ($p_{bb}$,$p_{bc}$) notation, where $p_{bb}$ is the polarization of the photon(s) that drive the bound-bound transition, where $p_{bc}$ is the polarization of the photon(s) that drive the bound-continuum transition and where $p_i=1$ for LCP, $p_i=-1$ for RCP and $p_i=0$ for linearly polarized photon(s). Note that for the all [1+1'] ionization schemes, the quantities $b_0$, $b_1$, $b_3$ and $MPPECD$ are approximated quantities ($\tilde{b_0}$, $\tilde{b_1}$, $\tilde{b_3}$ and $\widetilde{MPPECD}$) since the cylindrical symmetry of the interaction is broken.}
  \label{tab1}
  
\end{table*}

The simultaneous use of linearly and circularly polarized radiation breaks the cylindrical symmetry which is necessary to retrieve the photoelectron angular distribution from its projection on the velocity map imaging spectrometer by Legendre polynomials decomposition. Additional terms appear in the decomposition, which cannot be unequivocally determined from the projected images. Applying a conventional pBasex analysis can thus result in errors in the relative weights of the different Legendre coefficients. In order to extract the PAD and PECD we would need to directly resolve the 3D photoelectron distribution using another type of detector \cite{doerner00}, or to perform a tomographic reconstruction by recording projections of the PAD for different polarization directions of the linearly polarized pulses \cite{wollenhaupt09}. Nevertheless, we still decompose the projected images into Legendre polynomials to extract approximate distributions $\tilde{b_i}$, $\widetilde{PES}$ and $\widetilde{PECD}$. They may slightly differ from the actual 3D distributions, but these differences should not affect the conclusions drawn below, which are based on the overall shape of the distributions and on forward/backward integrated signals.  

Figure \ref{fig4}(a) shows the results obtained by driving the bound-bound transition with a linearly polarized ($p=0$) 199 nm photon and photoionizing with a circularly polarized ($p=\pm1$) 402 nm photon. The overall $\widetilde{PECD}$ is stronger than in the fully circular [2+1] and [1+1'] cases (see table \ref{tab1}). The $\widetilde{PECD}$ from the higher energy peak, associated to ionization from the HOMO, has the same sign as in the [1+1'] case and is about twice larger when driven by linear photons compared to circular. These results illustrate once again the importance of the photoexcitation anisotropy, and provide interesting information on the angular dependence of $\widetilde{PECD}$. For the HOMO, ionization of molecules selected by one or two-photon circular excitation produces  $\widetilde{PECD}$ with opposite signs. Because the bound-bound transition preferentially promotes the molecules that have their transition dipole parallel to the laser electric field into Rydberg states, the use of linear exciting photons photoselects an \textit{axis}, while the use of circularly polarized photon photoselects a \textit{plane} of molecular orientations. The PAD and the PECD being the incoherent sum over all the photoexcited molecular orientations, it is not surprising to see a difference when driving the bound-bound transition with linear or circular photons. Looking at table \ref{tab1}, it is also clear that the use of linear photon leads to a greater enhancement on the $b_3$ than on the $b_1$, since $b_3$ is the observable with the larger sensitivity to the photoexcitation anisotropy \cite{comby16,goetz17}.  

In order to determine the possible influence of the helicity of the exciting photons on the outcome of the PECD measurements, we performed additional experiments in which we used circularly polarized excitation pulses of 199 nm, associated to different polarization states of the ionizing pulses. First, we measured the forward-backward asymmetry obtained by photoexcitating limonene with 199 nm circularly polarized light (helicity $p=\pm1$) and ionizing with linearly polarized 402 nm pulses ($p=0$). 

A clear forward-backward asymmetry is observed, consisting solely of first order Legendre polynomials (pure $\tilde{b_1}$). In a recent experiment, we have demonstrated a new chiroptical effect called PhotoeXcitation Circular Dichroism, which explains the non-zero forward-backward asymmetry in this polarization configuration. A general analytical theory shows that PXCD originates from the superposition of intermediate states with non-collinear dipoles of transitions driven by circularly polarized light \cite{beaulieu18}. 

The importance of the photoexcitation helicity relative to the ionization step in PECD can be determined by comparing the PECD obtained from molecules photoexcited by left and right photons (Fig. \ref{fig4}(d,e,f)). We drive the bound-bound transition with a given circular polarization of the 199 nm pulse ($p=+1$ in (d) and $p=-1$ in (e)) and ionize alternatively with $p=+1$ and $p=-1$ circularly polarized 402 nm pulse in order to measure the difference between the PAD driven with co-rotating and counter-rotating two-color fields. Both cases lead to very similar PADs and PECDs. This observation strengthens the conclusion that the main role of the bound-bound transition is to photoselect some given molecular orientations (an axis for $p=0$, or a plane for $p=\pm1$). 

A synthetic view of the different cases studied is given in Table \ref{tab1}, which provides the values of the odd Legendre coefficients associated to the dominant ionization channels. These coefficients are normalized by the PES to enable direct quantitative comparison. 

In conclusion, we have studied the role of resonances in the multiphoton PECD of limonene. Using two-color fields in a fully tunable polarization scheme, we were able to disentangle the role of bound-bound and bound-continuum transitions. We have demonstrated experimentally that the bound-bound transitions do not need to be driven by circular photons in order to observe PECD, as recently theoretically predicted \cite{goetz17}. The helicity of the pulse that drives the bound-bound transitions hardly modifies the measured PECD. However circular and linear photoexciting photons still lead to different PECDs, because of the angular dependence of the excitation process which induces different anisotropies. These results illustrate the importance of molecular alignment in PECD experiments, and call for the systematic investigation of PECD in aligned molecules, through electron-ion coincidence imaging \cite{tia17} or using laser-induced molecular alignment \cite{roscapruna01,christensen15}.

We thank R. Bouillaud and L. Merzeau for technical assistance and Bernard Pons for fruitful discussions. This project has received funding
from the European Research Council (ERC) under the European Union's Horizon 2020 research and innovation programme no. 682978 - EXCITERS. We acknowledge the financial support of the French National Research Agency through ANR-14-CE32-0014 MISFITS and IdEx Bordeaux – LAPHIA (ANR-10-IDEX-03-02). S.B. acknowledge the NSERC Vanier Scholarship.


%

\end{document}